\input{aipcheck}

\documentclass[
    ,final            % use final for the camera ready runs
  ]
  {aipproc}

\layoutstyle{8x11single}

\begin{document}

\title{Diffractive Measurements in ATLAS}

%\Keywords{Diffraction, Rapidity gaps, Cross-sections}

\author{K. Shaw (on behalf of the ATLAS collaboration)}{
  address={INFN Gruppo Collegato di Udine and ICTP, Italy}
}

%\author{<author2>}{
%  address={<common address for author2 and author3>}
%}

%\author{<author3>}{
%  address={<common address for author2 and author3>}
%  ,altaddress={<author1 address>} % additional visiting address
%}

\begin{abstract}
Measurements made using the ATLAS detector at the LHC at $\sqrt{s}$~=~7~TeV incorporating diffractive processes are presented. A first measurement of the inelastic cross-section using~20~$\mu$b$^{-1}$ of data is given, yielding a result of ~$\sigma_{inel}(\xi > $~5~$\times$~10$^{-6})$~=~60.3~$\pm$~2.1~mb, for single ($pp \rightarrow Xp$) and double ($pp \rightarrow XY$) diffractive processes for a kinematic range corresponding to detector acceptance~$\xi$~=~M$^2_X/s$ calculated from the invariant mass~M$_X$ of the heavier dissociation system X. %dissociative system.
 Furthermore a study is made of pseudorapidity gap distributions using~7.1~$\pm$~0.2~$\mu$b$^{-1}$ of data collected to tune the diffractive fraction of the inelastic cross-section in Monte Carlo (MC) models, and a measurement is made of the differential cross-section for events with large gaps in pseudorapidity where diffractive processes dominate.

\end{abstract}

\maketitle

%%%%%%%%%%%%%%%%%%%%%%%%%%%%%%%%%%%%%%%%%%%%
%% MAINMATTER
%%%%%%%%%%%%%%%%%%%%%%%%%%%%%%%%%%%%%%%%%%%%

\section{Introduction}

Diffractive processes are of interest to study as they are the dominant contribution to high-energy elastic and quasi-elastic scattering in hadron interactions and they contribute to the total cross-section which cannot be calculated directly from Quantum Chromodynamics (QCD). Moreover in the high particle multiplicity environment of the Large Hadron Collider (LHC), it is of great importance to describe the large number of additional \textit{pp} interactions in each bunch crossing, which has a significant contribution from diffractive dissociation. Thus using the ATLAS detector \cite{atlas2008} at the LHC a first measurement has been made of the inelastic cross-section \cite{inelastic2011} and a study of pseudorapidity gap\footnote{Known as simply `rapidity gaps' for the remainder of the note.} signatures of diffractive events \cite{rapiditygap2011} has been performed resulting in a first measurement of the differential cross-section. % at the LHC are presented. %These measurements cannot be calculated from QCD, although alternative theories provide some description, thus experimentally derived measurements are vital to develop a full understanding of these processes. 

The total \textit{\textit{pp}} cross-section at the LHC can be divided into elastic (el) and inelastic (inel) components. %In elastic events there is no colour flow between colliding partons, 
The inelastic part is further divided into single diffractive (sd), double diffractive (dd), central diffractive (cd) and non-diffractive (nd) components, $\sigma_{tot} = \sigma_{el} + \sigma_{sd}+ \sigma_{dd} + \sigma_{cd} + \sigma_{nd}$. Only non-diffractive inelastic events involve a colour exchange between the colliding partons. %where the probability of large momentum transfer is high and thus these interactions produce a uniform distribution of debris. 
In diffractive events there is a colour singlet exchange, which can be described using Regge theory \cite{regge1} as a Pomeron, resulting in large gaps in pseudorapidity\footnote{Pseudorapdity is defined in terms of the polar angle $\theta$ as $\eta$ = -ln tan ($\theta$ / 2).} ($\eta$) between the proton dissociation products, where the debris is produced in the very forward regions of the detector.

For these measurements Minimum Bias Trigger Scintillators (MBTS) were used to trigger on activity on only one side (single-sided) or both sides (inclusive) of the interaction point (IP), and consist of two sets of sixteen scintillator counters situated on the inner face of the endcap tile calorimeter in the forward regions of ATLAS. %, and can trigger on particle activity on only one side (single-sided) or both sides (inclusive) of the interaction point (IP).  % at \textit{z} = $\pm$3.6m from the interaction point {IP}. 
 Additionally the ATLAS inner detector and calorimeters provide information on particle activity distributions in pseudorapidity.

\section{Measurement of  the Inelastic Cross-section}

Using an integrated luminosity of 20.3 $\pm$ 0.7 $\mu$b$^{-1}$ at $\sqrt{s}$ = 7 TeV in 2010 the inelastic \textit{\textit{pp}} cross-section has been measured using an inclusive event sample selecting events where at least two MBTS scintillating counters detect a charge larger than some threshold, known as hits. Additionally to constrain the diffractive component of the inclusive sample, single-sided events were selected where one side of the MBTS detector contained at least two hits, and the other side detected no hits. 

The measurement is limited to events where the dissociation systems have a large invariant mass (M$_X$, where X denotes the larger (only) dissociative mass system separated by a rapidity gap with the other dissociative mass system (proton) in double (single) diffractive events), as these are outside the acceptance of the MBTS corresponding to $\xi = $ M$^2_X / s > $ 5 $\times$ 10$^{-6}$, equivalent to M$_X >$ 15.7 GeV. Thus the inelastic cross-section has been measured according to Equation~\ref{eq:crosssec}, where N (N$_{BG}$) denotes the number of selected (background) events within the MBTS acceptance, $\int Ldt$ is the integrated luminosity, $\epsilon_{trig}$ ($\epsilon_{sel}$) is the trigger (selection) efficiency in the $\xi$-range and $f_{\xi < 5 \times 10^{-6}}$ is the fraction of events that pass the event selection with $\xi < 5 \times 10^{-6}$.

\begin{equation}\label{eq:crosssec}
\sigma (\xi > 5 \times 10^{-6}) = \frac{N - N_{BG}} {\epsilon_{trig} \times \int Ldt} \frac{1 - f_{\xi < 5 \times 10^{-6}}} {\epsilon_{sel}}
\end{equation}

\begin{figure}\label{fig:ratio}
  \includegraphics[height=.3\textheight]{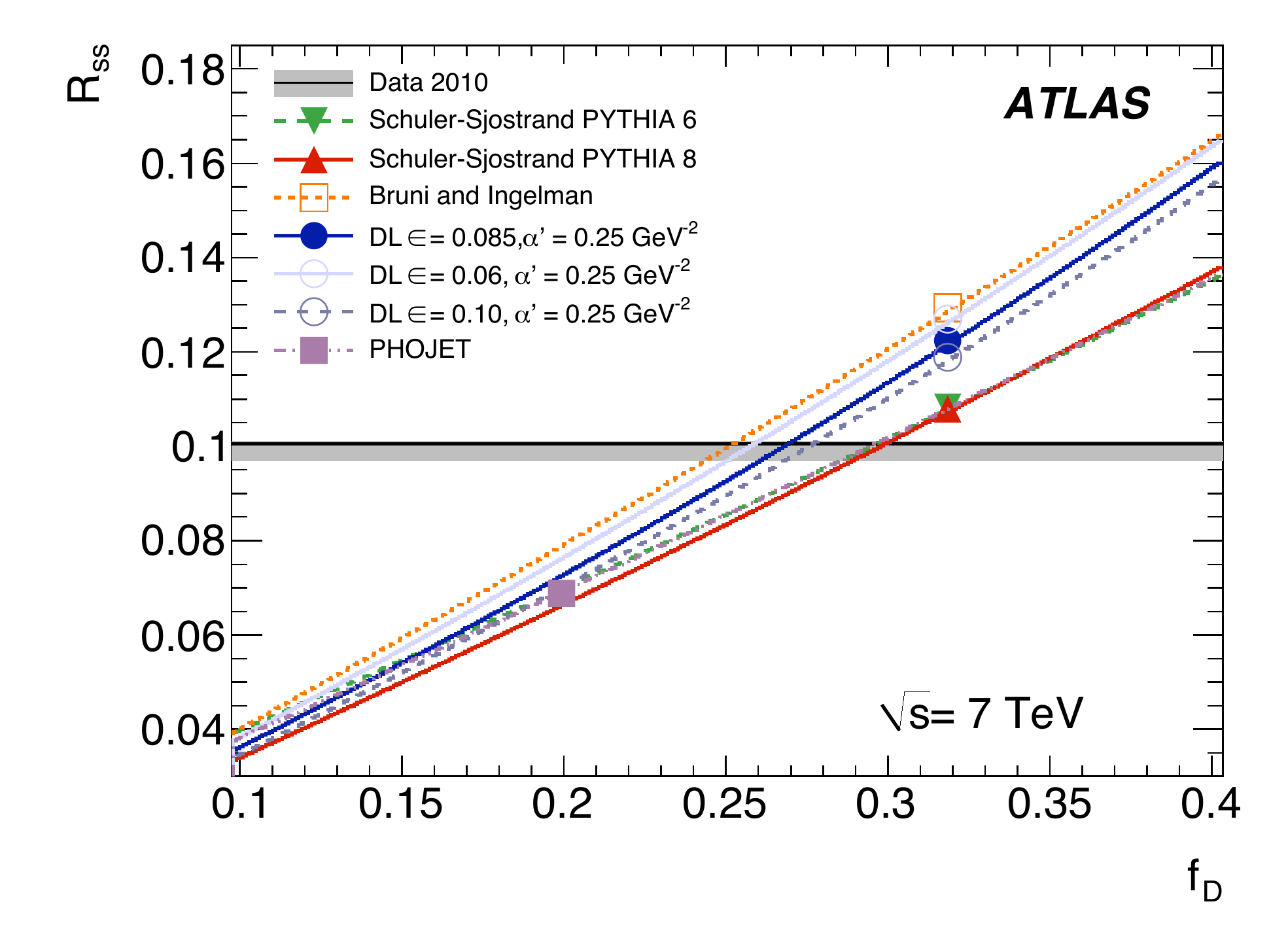}
  \includegraphics[height=.3\textheight]{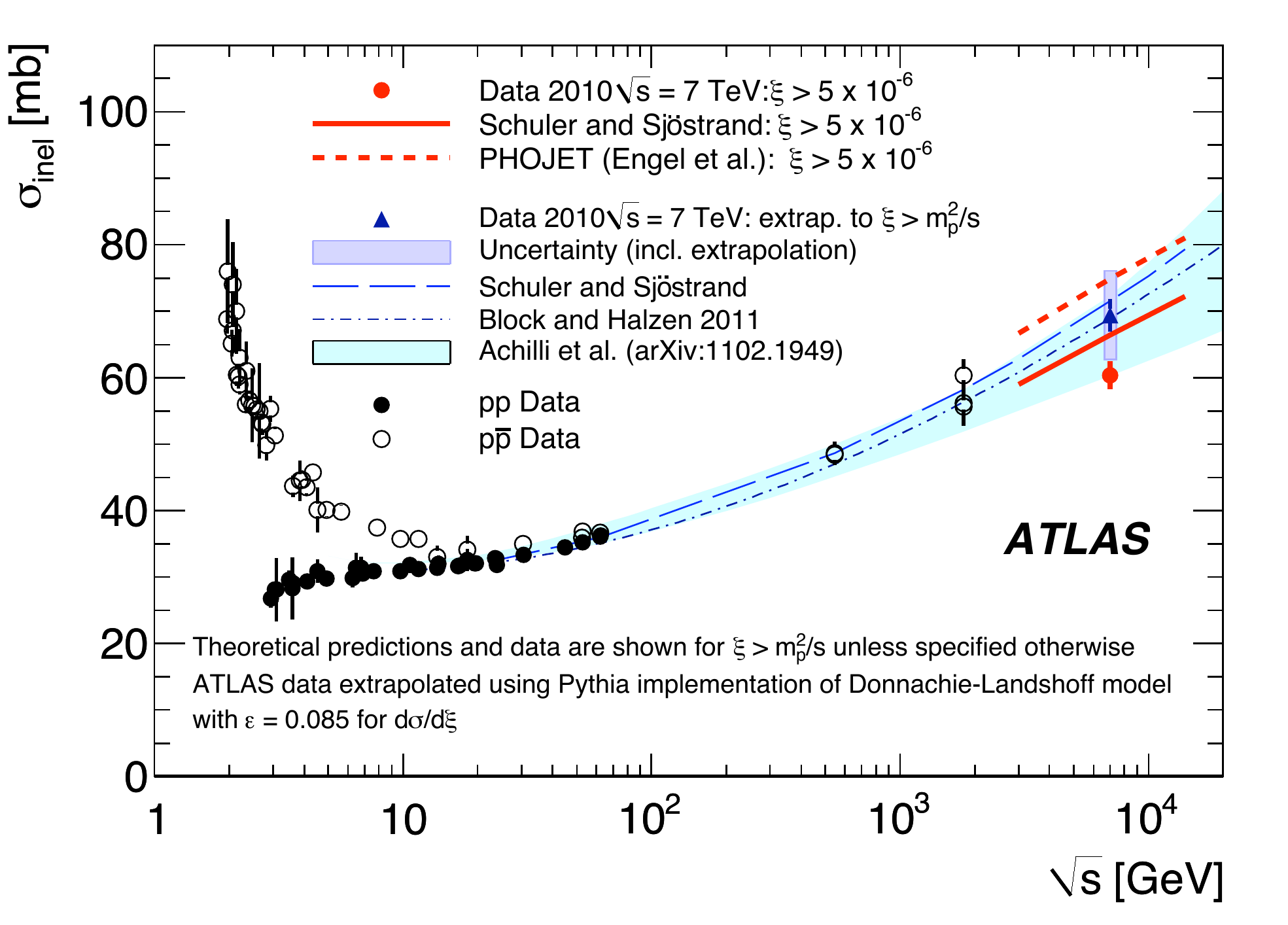}
  \caption{The left-hand plot shows the uncorrected ratio R$_{ss}$ of single-sided events, dominated by diffractive processes, to inclusive events, dominated by non-diffractive processes, as a function of the fractional contribution of diffractive events to the inclusive sample (f$_D$). The right-hand plot shows the inelastic cross-section as a function of $\sqrt{s}$ where the ATLAS measurement for $\xi > 5 \times 10^{-6}$ is shown (red filled circle) and the measurement extrapolated to the full inelastic cross-section is shown (blue triangle) with error bars indicating the associated uncertainty including the extrapolation uncertainty (blue shaded area).}
\end{figure}

The fractional contribution of diffractive events (f$_D = \sigma_{sd} + \sigma_{dd} / \sigma_{inel}$) to the total inelastic cross-section varies significantly between models. Thus the ratio of single-sided to inclusive events ($R_{ss}$) in data, measured to be R$_{ss,(data)}$~=~10.02~$\pm$~0.03(stat.)~$^{+0.1}_{-0.4}$(syst.)$\%$, is sensitive to f$_D$ in the MC models and therefore can be used to constrain f$_D$. The left-hand plot in Figure \ref{fig:ratio} compares R$_{ss,(data)}$ to  R$_{ss}$ predicted from a variety of MC generators with varying values of f$_D$, where their intersection with the R$_{ss,(data)}$ is used as the central value for f$_D$ for that model. Using the default DL~model~\cite{dl} the resulting value is f$_D$ = 26.9$^{2.5}_{-1.0} \%$ and this tuned model is used to calculate MC dependent corrections to the cross-section measurement.  

The inelastic cross-section has been measured as $\sigma_{inel}(\xi > 5 \times 10^{-6})$~=~60.3~$\pm$~0.05(stat.)~$\pm$~0.5(syst.)~$\pm$~2.1~(lumi)~mb. Furthermore to compare this with previous results the measurement has been extrapolated to the full inelastic cross-section giving  $\sigma_{inel}(\xi > m_p^2 /s $)~=~69.4~$\pm$~2.4~(exp.)~$\pm$~6.9~(extr.)~mb and is shown on the right-hand plot in Figure~\ref{fig:ratio} with the kinematically constrained measurement and compared with several MC predictions.

\section{Differential Cross-Section Measurements For Rapidity Gaps}
The diffractive contribution to the inelastic cross-section has been tuned using 7.1 $\pm$ 0.2 $\mu$b$^{-1}$ of data and the differential cross-section has been measured.
%of data has been used to tune the diffractive contribution to the inelastic cross-section, and measure the differential cross-section. %The cross-section is defined here in terms of stable particles with transverse momentum, $p_T$ larger than a threshold $p_T^{cut}$, referred to as 'hadron level'. 
The measurement of the masses of dissociative system(s) is complicated by the fact that much of the debris is produced beyond the acceptance of the detector. A method has been developed to obtain the cross-section as a function of the size of the visible part of the rapidity gap separating the two systems (X and Y) in the case of double-diffractive processes, or separating the dissociative system from the elastically scattered proton in the case of single-diffractive processes. 
Rapidity gaps are identified by dividing the ATLAS calorimeters and inner detector into rings in $\eta$ and identifying the largest sequential runs of rings without any particle activity in them, known as empty rings, starting from the edge of the acceptance( $\eta$ = $\pm$ 4.9) and of size ($\Delta \eta^F$) . Particle activity is defined in the calorimeter as a cell above a chosen threshold, and in the tracker as a reconstructed track where the $p_T >$ 200 MeV, $|\eta| < $ 2.5. 
%add in here aout the pt?200 thing
MC generators such as Pythia~\cite{pythia} and Phojet~\cite{phojet} have varying values of f$_D$. Thus these fractions were optimised for each model by fitting their diffractive and non-diffractive components to detector level data. This optimized fraction was found for the default MC generator Pythia8 to be f$_D$ = 30.2 $\pm$ 0.3(stat.) $\pm$ 3.8(syst.)$\%$. %This is a nice result as it is consistent with the value for $R_{ss}$ found from the study shown in the previous section.

% have different default diffractive fractions of the total inelastic cross-section, especially for the double diffractive component. These fractions were optimised for each MC model, by fitting their diffractive and non-diffractive components to detector level data. The diffractive fraction for the default MC generator Pythia8 after optimisation has been measured to be f$_D$ = 30.2 $\pm$ 0.3(stat.) $\pm$ 3.8(syst.)$\%$. This is a nice result as it is consistent with the value for $R_{ss}$ found from the study shown in the previous section.

\begin{figure}\label{fig:results}
  \includegraphics[height=.3\textheight]{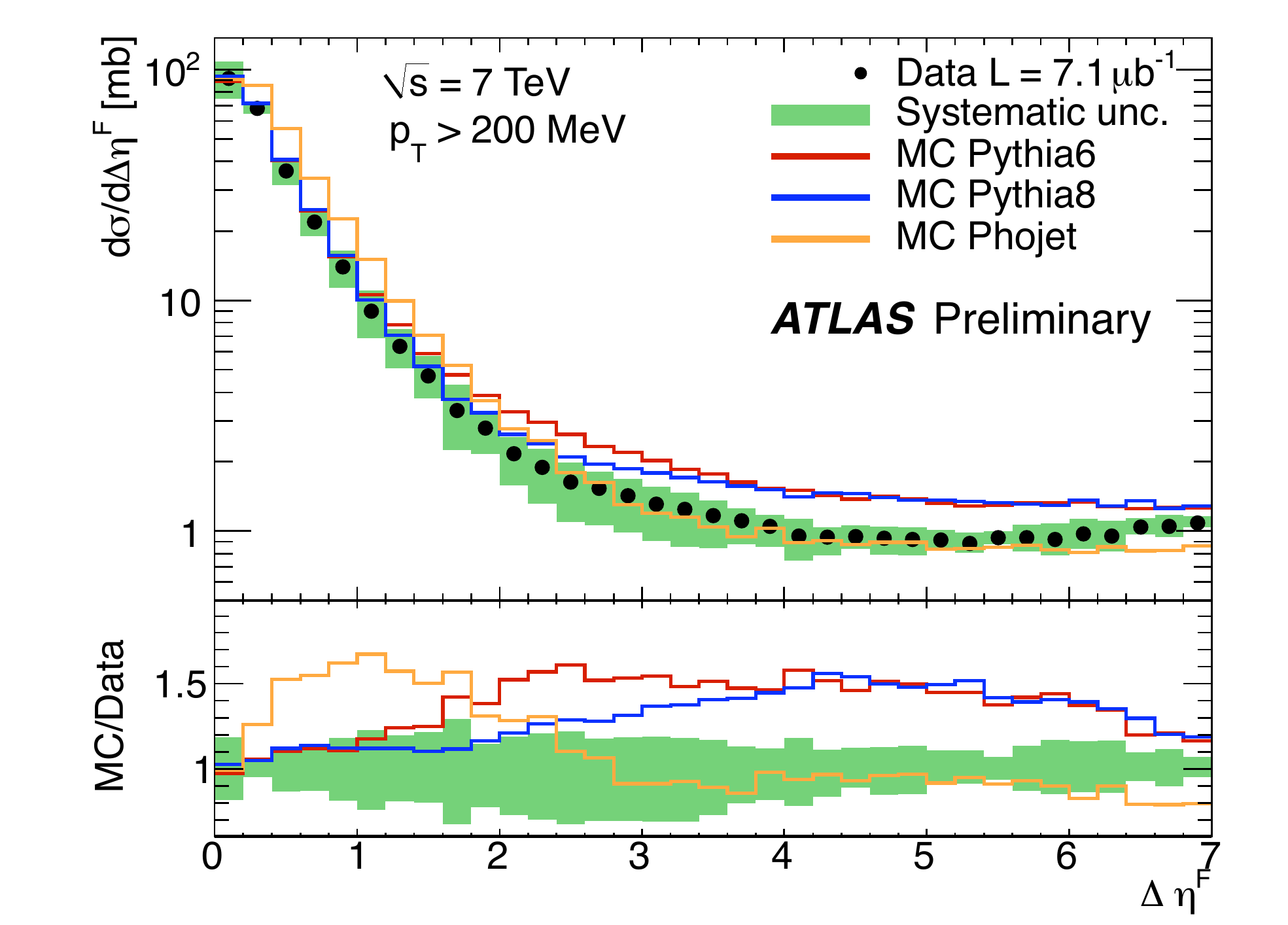}
  \includegraphics[height=.3\textheight]{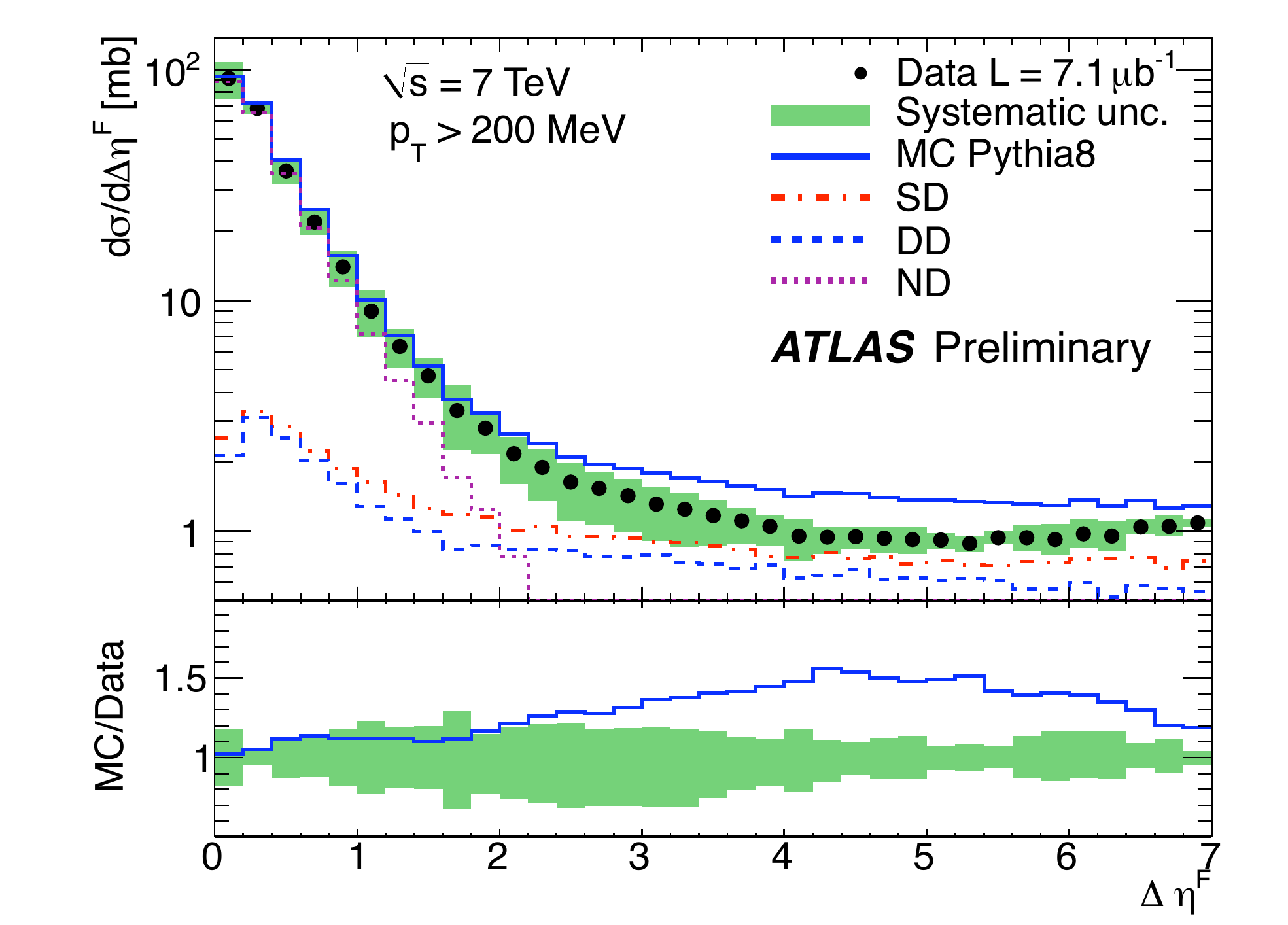}
  \caption{ Inelastic cross-section shown as a function of rapidity gap size ($\Delta \eta^F$) for particles with $p_T >$ 200 MeV corrected for experimental effects. Left-hand plot shows a comparison between various MC models and data and the right-hand plot shows the cross-section for only the default Pythia8 MC generator, with the single, double and non-diffractive contributions shown separately.}
\end{figure}

%The observable measured is  where $\Delta \eta^F$ is the size of the rapidity gap in the forward region. 
The left-hand plot of Figure \ref{fig:results} shows the measured differential inelastic cross-section ($d\sigma / d \Delta \eta^F$) compared to various MC generators which shows at small (large) $\Delta \eta^F$ the Pythia (Phojet) MC describes the data best. The right-hand plot presents the different contributions to the cross-section for Pythia 8 MC generator showing at small (large) values of $\Delta \eta^F$  the non-diffractive (diffractive) component dominates. The diffractive cross-section has been measured as $d\sigma / d\Delta \eta^F \approx$ 1.0 $\pm$ 0.2 mb per unit of $\Delta \eta^F$ at high $\Delta \eta^F$ for $p_T$ > 200 MeV.  
   
% at small values of $\Delta \eta^F$ with the  components dominating at larger values $\Delta \eta^F$.
  
\section{Conclusion}

A first measurement of the inelastic cross-section at $\sqrt{s}$ = 7 TeV has been presented limited to the kinematic range corresponding to the acceptance of the ATLAS detector $\xi > 5 \times 10^{-6}$. Additionally the inelastic cross-section has been measured as a function of forward rapidity gaps.% finding an exponentially falling non-diffractive component in the low $\Delta \eta^F$ regions and a plateau of single and double diffractive events which is well described in shape by MC models. 

%\begin{figure}
%  \includegraphics[height=.3\textheight]{golfer}
%  \caption{Picture to fixed height}
%\end{figure}

%\begin{table}
%\begin{tabular}{lrrrr}
%\hline
%  & \tablehead{1}{r}{b}{Single\\outlet}
%  & \tablehead{1}{r}{b}{Small\tablenote{2-9 retail outlets}\\multiple}
%  & \tablehead{1}{r}{b}{Large\\multiple}
%  & \tablehead{1}{r}{b}{Total}   \\
%\hline
%1982 & 98 & 129 & 620    & 847\\
%1987 & 138 & 176 & 1000  & 1314\\
%1991 & 173 & 248 & 1230  & 1651\\
%1998\tablenote{predicted} & 200 & 300 & 1500  & 2000\\
%\hline
%\end{tabular}
%\caption{Average turnover per shop: by type
%  of retail organisation}
%\label{tab:a}
%\end{table}

%\end{theacknowledgments}

%%%%%%%%%%%%%%%%%%%%%%%%%%%%%%%%%%%%%%%%%%%%
\bibliographystyle{aipproc}   % if natbib is available
%\bibliographystyle{aipprocl} % if natbib is missing

%%%%%%%%%%%%%%%%%%%%%%%%%%%%%%%%%%%%%%%%%%%
%% You probably want to use your own bibtex database here
%%%%%%%%%%%%%%%%%%%%%%%%%%%%%%%%%%%%%%%%%%%
%\bibliography{sample}

%%%%%%%%%%%%%%%%%%%%%%%%%%%%%%%%%%%%%%%%%%%
%% Just a reminder that you may have to run bibtex
%% All of it up to \end{document} can be removed
%% if you don't like the warning.
%%%%%%%%%%%%%%%%%%%%%%%%%%%%%%%%%%%%%%%%%%%
\IfFileExists{\jobname.bbl}{}
 {\typeout{}
  \typeout{******************************************}
  \typeout{** Please run "bibtex \jobname" to optain}
  \typeout{** the bibliography and then re-run LaTeX}
  \typeout{** twice to fix the references!}
  \typeout{******************************************}
  \typeout{}
 }
\footskip = -400pt
\let\thefootnote\relax\footnotetext{Copyright \copyright  \hspace{0.5mm}  CERN for the benefit of the ATLAS Collaboration}

%%%%%%%%%%%%%%%%%%%%%%%%%%%%%%%%%%%%%%%%%%%
%% The following lines show an example how to produce a bibliography
%% without the help of the BibTeX program. This could be used instead
%% of the above.
%%%%%%%%%%%%%%%%%%%%%%%%%%%%%%%%%%%%%%%%%%%

%\endinput
%%
%% End of file `template-8s.tex'.
\end{document}